\documentclass[twoside,fleqn]{article}
\usepackage{epsf}
\usepackage{espcrc2}
\newcommand{\disable}[1]{}

\newenvironment{myeqnarray}{\arraycolsep 0.14em\begin{eqnarray}}{\end{eqnarray}}

\newcommand{\journal}[5]{#1;\textsl{ #2} \textbf{#3 }(#4) #5}
\newcommand{\ba}{\begin{myeqnarray}}
\newcommand{\ea}{\end{myeqnarray}}
\newcommand{\be}{\begin{equation}}
\newcommand{\ee}{\end{equation}}

\newcommand{\ccbar}{c\bar{c}}
\newcommand{\bbbar}{b\bar{b}}
\newcommand{\mathbi}[1]{\mbox{\boldmath $#1$}}
\newcommand{\p}{\mathbi{p}}
\newcommand{\psb}{\bar{\psi}}
\newcommand{\oa}{\mathcal{O}(a)}

\newcommand{\fij}{F_{ij}}
\newcommand{\foi}{F_{0i}}

\newcommand{\onepicture}[1]{
\begin{picture}(2.5,2.05)
  \put(.04,-.42){\epsfxsize=2.7in \epsfbox[10 30 560 590] {./#1}}
\end{picture}
}

\newcommand{\twopictures}[2]{
\begin{picture}(2.6,3.65)
\put(.03,1.5){\begin{picture}(2.5,2.1)
    \put(0,0){\epsfxsize=2.8in \epsfbox[10 30 560 590]{./#1}}
  \end{picture}}
\put(.03,-.5){\begin{picture}(2.5,2.1)
    \put(0,0){\epsfxsize=2.8in \epsfbox[10 30 560 590]{./#2}}
  \end{picture}}
\end{picture}
}

\title{First results from the asymmetric $\oa$ improved Fermilab action}
\author{Z. Sroczynski
\address{Department of Physics, University of Illinois, 1110 West
Green Street, Urbana, IL 61801, USA}
}

\begin{document}

\begin{abstract}
We present first results from calculations using an $\oa$
improved (FNAL) space-time asymmetric fermion action on a
$12^3\times 24$ quenched lattice at $\beta=5.7$ and with $c_\mathrm{SW}=1.57$.
The mass dependent asymmetry parameter $\zeta$ is determined
non-perturbatively from the energy-momentum dispersion relation.
 Calculations have been
made in the charm and bottom quark mass sectors in order to test the
$\zeta$ dependence of the spectrum, since it is at these
heavier masses that the asymmetry is expected to be most relevant.
\end{abstract}

\maketitle

\section{The Fermilab improved action}
For full details of this fermionic improvement
scheme the reader is referred to \cite{action}; 
here we merely note that it results in an action with mass dependent
coefficients which is asymmetric in space and time.

The lattice dispersion relation may be written in the form
\be\label{dispersion}
E^2(\p^2) = M_1^2+\frac{M_1}{M_2}\p^2 + \mathcal{O}(\p^4), 
\ee
defining the {static mass} $M_1 = E(\mathbf{0})$ and the
\mbox{kinetic mass} 
$M_2 = \left(\frac{\partial^2E}{\partial p_i^2}\right)^{-1}_{\!\p=\mathbf{0}}$. 
\begin{figure}[h]
\onepicture{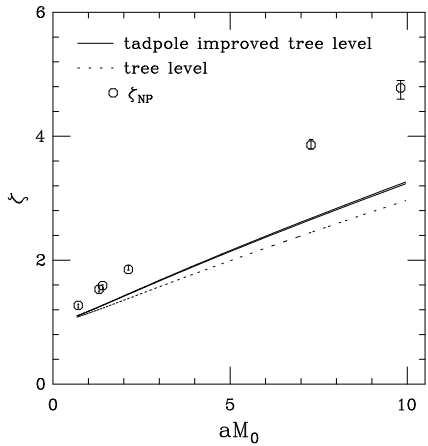}
\caption{\label{zeta_m0}The non-perturbatively tuned
$\zeta$ compared with tree-level perturbative predictions.}
\end{figure}

Lattice discretisation effects mean than in general $M_1\neq M_2$ at
$\mathcal{O}(am_\mathrm{quark})$, and restoring the relativistic dispertion
relation to this order is 
the first stage in the improvement program. This is
achieved by introducing an asymmetry in the temporal and spatial
quark propagation and adjusting it until $M_1\!=\!M_2$, which then
constitutes the first improvement condition of this scheme. 

To this end we define
the action 
\ba\label{hopping}
S_0& = &\sum_x\big\{\psb_x\psi_x\nonumber\\
&&-\kappa_t\left[\right.\psb_x(1\!-\!\gamma_0)U_{0\,x}\psi_{x+\hat{0}}\nonumber\\
\rule{0pt}{14pt}&&\hspace{1.7em}+\psb_x(1\!+\!\gamma_0)U^{\dagger}_{0\,x-\hat{0}}\psi_{x-\hat{0}}\left.\right]\nonumber\\
\rule{0pt}{14pt}&&-\kappa_s\sum_i\left[\right.\psb_x(1\!-\!\gamma_i)U_{i\,x}\psi_{x+\hat{i}}\nonumber\\
&&\hspace{3.3em}+\psb_x(1\!+\!\gamma_i)U^{\dagger}_{i\,x-\hat{i}}\psi_{x-\hat{i}}\left.\right]\big\}.
\ea
It is helpful to parameterise this asymmetry by defining
$\zeta=\kappa_s/\kappa_t$, in terms of which the quark mass is
\be\label{m0def}
M_0 = \frac{1}{2\kappa_t}-3\zeta -1 - \left(\frac{1}{2\kappa_\mathrm{crit}}-4\right).
\ee
At some value $\zeta\!=\!\zeta_\mathrm{NP}$, which we attempt here to
find at various quark masses, $M_1\!=\!M_2$.

In order to remove $\oa$ artifacts from the action the terms

\be S_E=i\kappa_sc_E\sum_{x,i}\psb_x\sigma_{0i}\foi(x)\psi_x \ee
\vspace{-0.1in}and
\be S_B=i\kappa_sc_B\sum_{x,i<j}\psb_x\sigma_{ij}\fij(x)\psi_x \ee
are added to $S_0$. Here $\foi$ and $\fij$ are the standard clover
representations of chromo-electric and  chromo-magnetic parts of the
field strength tensor.

\section{Non-perturbative tuning of $\zeta$}

\begin{figure}
\onepicture{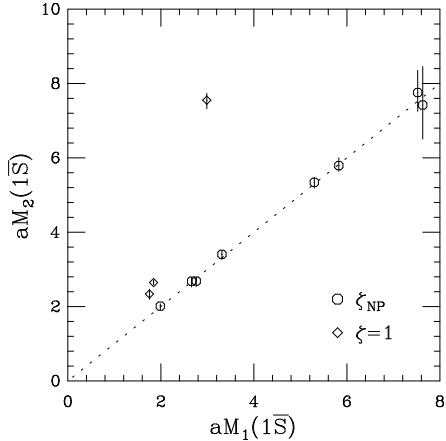}
\caption{\label{M1_M2}$M_2$ and $M_1$ at 
$\zeta_\mathrm{NP}$ compared with the $M_1\!=\!M_2$ line and the
corresponding results from the symmetric action.}
\end{figure}

The strategy employed here was to tune $\zeta$ by requiring that
$M_1\!=\!M_2$ for the spin-averaged $1S$ quarkonium state. 
The results obtained from this non-perturbative tuning can be compared
where appropriate with tree-level perturbation theory \cite{action,pt}
and with results obtained using the symmetric  ($\zeta\!=\!1$) action \cite{symmetric}.  

These calculations were performed on 100 quenched $12^3\times 24$
configurations at $\beta=5.7$ with
$c_E=c_B=1.57$,  the tree-level tadpole
improved  perturbative value on this lattice.

To find $M_1$ and $M_2$, the ground state energy $E(\p)$ 
 was computed at five momenta using a two-state fit to a
matrix of smeared correlators (as described in \cite{multistate}).
$M_1$ is simply $E(\mathbf{0})$ 
and $M_2$ was extracted from the coefficient $a_1$ obtained by fitting
the dispersion relation to the function
\be
E(\p^2) = a_0+a_1\p^2+a_2(\p^2)^2+a_3\sum_ip_i^4.
\ee

We obtain a graph (figure \ref{zeta_m0}) of the non-perturbatively
tuned $\zeta_\mathrm{NP}$ as a function of $M_0$ which we compare with
tree level perturbation theory. 
The extent to which  $\zeta_\mathrm{NP}$ satisfies the improvement
condition  can be
judged from figure \ref{M1_M2}, where for comparison the corresponding
points previously obtained using the symmetric action on
the same lattice are also plotted. 

\begin{figure}
\onepicture{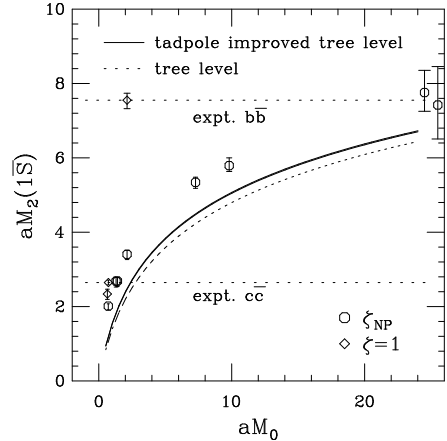}
\caption{\label{M2_m0}The variation of $M_2$ with $M_0$ for the tuned
action, compared with the symmetric action and tree-level perturbation theory.}
\end{figure}

While figure \ref{zeta_m0}  presents the dependence of
$\zeta_\mathrm{NP}$ upon $M_0$, it is $M_2$ that emerges as the
physically significant mass parameter in the heavy quark expansion.
Therefore it is useful to know how $M_2$ depends on $M_0$ once $\zeta$
has been tuned, and this is shown in figure \ref{M2_m0}.

\section{Quarkonium spectrum}

\begin{figure}[t]
\twopictures{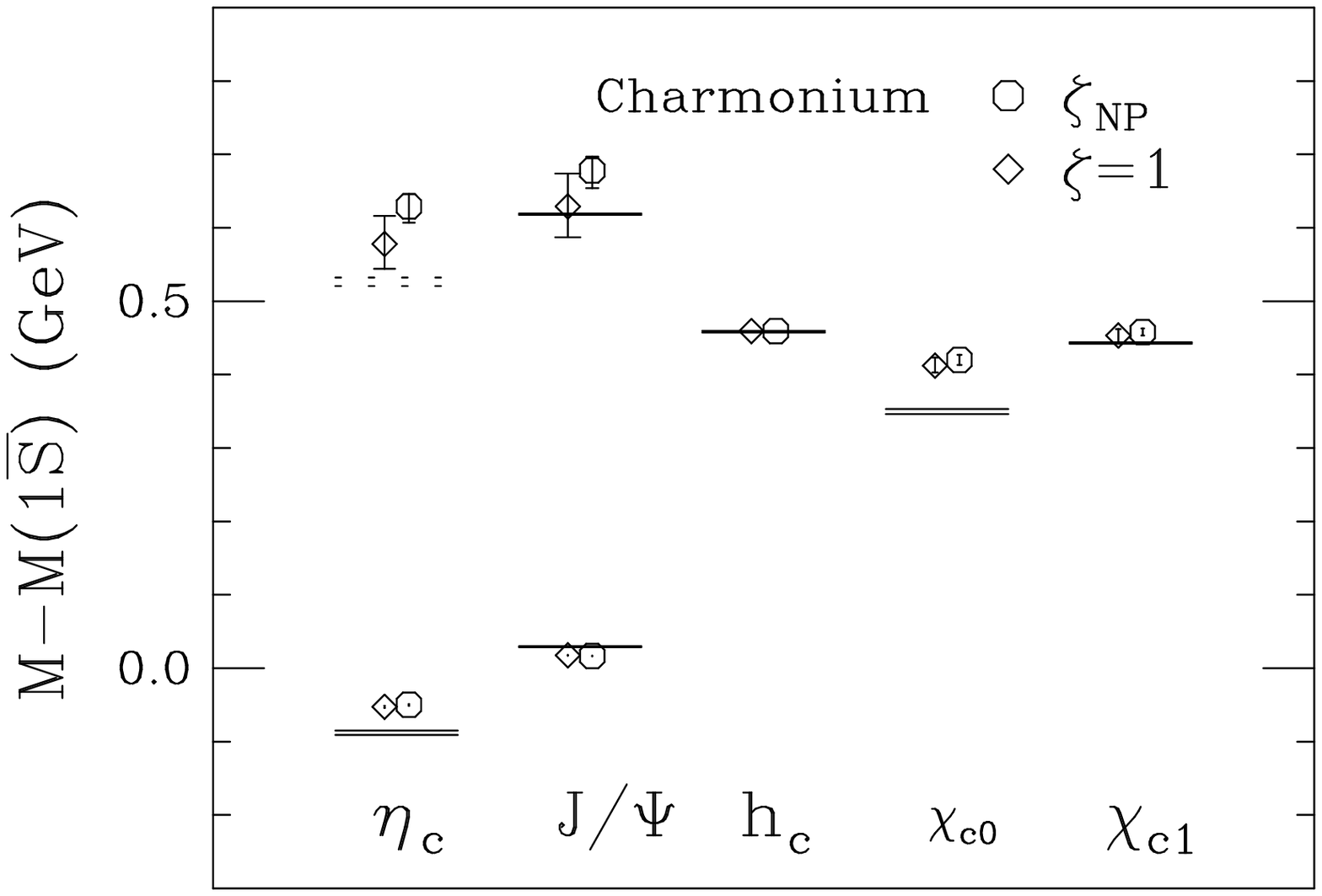}{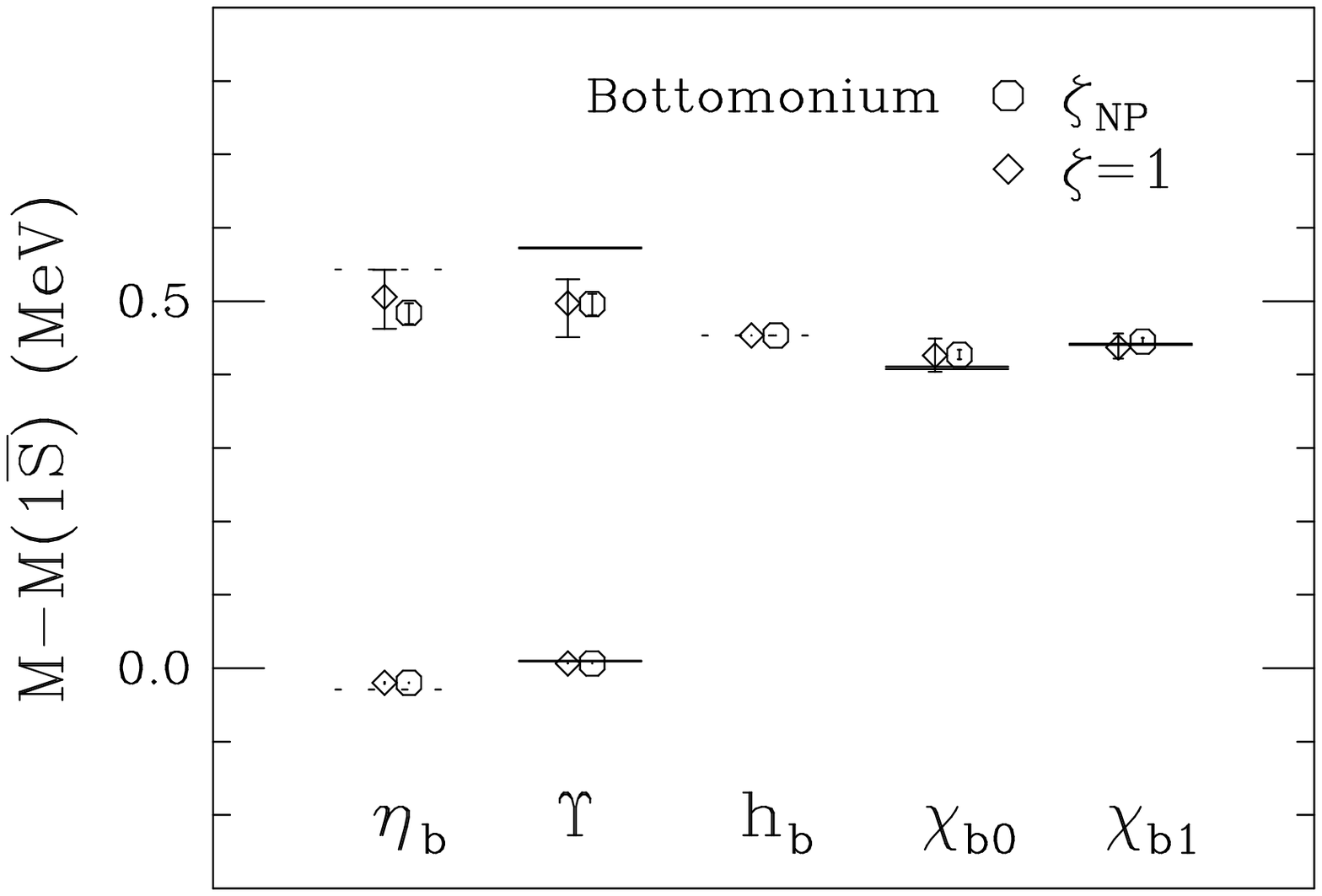}
\caption{\label{spectrum}The quarkonium spectra
compared with results obtained with the symmetric action.}
\end{figure}

\begin{figure}[t]
\onepicture{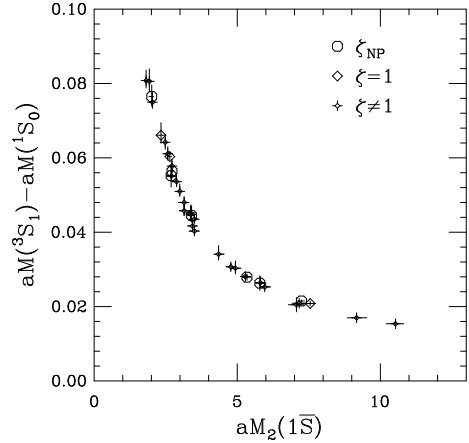}
\caption{\label{hyperfine_M2}The hyperfine splitting as a function of $M_2$ 
computed using the symmetric, the tuned and the asymmetric untuned actions.}
\end{figure}

The values of $aM_2$ that correspond on this lattice to the $\ccbar$
and $\bbbar$ mesons are known from the previous calculations with the
symmetric action \cite{symmetric}, and 
parameters of the asymmetric action yielding values of $aM_2$
comparable to these were found (see figures \ref{M2_m0} and \ref{zeta_m0}). 
To verify that the asymmetric action
reproduces the same physics a spectral calculation was performed at these
parameters on 300 configurations. $2S$ states were obtained using a
three-state fit to the full correlator matrix. The scale was set from
the spin-averaged $1P\!-\!1S$ splitting.
Figure \ref{spectrum} shows the masses of charmonium and bottomonium
states expressed as splittings from the spin-averaged $1S$ mass. 
The hyperfine splitting is shown in figure \ref{hyperfine_M2} as a
function of $M_2$ for all the data sets examined 
(\textit{i.e.} with untuned asymmetric actions as
well as at $\zeta_\mathrm{NP}$ and $\zeta=1$). 

\section{Conclusions}
We have demonstrated the feasibility of a non-perturbative tuning of
the first parameter of the Fermilab improved action, and find 
 this value to be rather higher than the tree level prediction,
although the mass dependence is already qualitatively predicted at
tree level. The bare
quark mass required to reach a particular physical r\'egime is found
to be much greater than for the symmetric action (again this
behaviour is  qualitatively predicted at tree level) which can have
algorithmic implications in the quark propagator computation. 

The quarkonium spectra (and a similar analysis of the hyperfine, fine
structure and the $2S\!-\!1S$ splittings) show that the tuned action
produces the expected 
physics, and supports the use of $M_2$ as the physically relevant mass
scale in computations where $\zeta$ is not tuned. Further confirmation
comes from figure \ref{hyperfine_M2} where it can be seen that the
hyperfine splittings lie on the same curve regardless of whether
the action is tuned or not, indicating that it is dependent on $M_2$ only.

\vfill

\section*{Acknowledgements}

This work was carried out in collaboration with Aida El-Khadra, Jim Simone, \mbox{Andreas} Kronfeld and Paul Mackenzie.

\disable{Ta\\ very\\ much.}

\end{document}